\documentclass[usenatbib]{mnras}
\usepackage{newtxtext,newtxmath}

\usepackage{graphicx}	

\usepackage[T1]{fontenc}
\usepackage{times}
\usepackage{url}
\usepackage{graphicx}
\usepackage{graphics} 
\usepackage{epsfig}   
\usepackage{microtype}
\usepackage{threeparttable} 
\usepackage{pdflscape}	
\usepackage{natbib}
\newcommand{\BAYMAX}{{\tt BAYMAX}}

\def\aj{{AJ}}                   
\def\araa{{ARA\&A}}          
\def\apj{{ApJ}}                 
\def\apjl{{ApJ}}                
\def\apjs{ {ApJS}}               
\def\ao{ {Appl.~Opt.}}           
             
\def\aap{ {A\&A}}                
\def\aapr{ {A\&A~Rev.}}

\def\mnras{ {MNRAS}}

\def\ssr{ {Space~Sci.~Rev.}}     
                 
\def\nat{ {Nature}}

\newcommand{\cxo}{\textit{Chandra}}

\newcommand{\xmm}{\textit{XMM}}

\newcommand{\be}{\begin{equation}}
\newcommand{\ee}{\end{equation}}

\newcommand{\gtsima}{$\; \buildrel > \over \sim \;$}
\newcommand{\ltsima}{$\; \buildrel < \over \sim \;$}
\newcommand{\prosima}{$\; \buildrel \propto \over \sim \;$}
\newcommand{\gsim}{\lower.5ex\hbox{\gtsima}}
\newcommand{\lsim}{\lower.5ex\hbox{\ltsima}}
\newcommand{\simgt}{\lower.5ex\hbox{\gtsima}}
\newcommand{\simlt}{\lower.5ex\hbox{\ltsima}}
\newcommand{\simpr}{\lower.5ex\hbox{\prosima}}

\newcommand{\so}{4C~63.20}
\newcommand{\bla}{blazar}

\newcommand{\hz}{{high-$z$}}
\newcommand{\ubrad}{$U_{\rm B}/U_{\rm rad}$}

\usepackage{graphicx}	
\usepackage{amsmath}	
\usepackage{amssymb}	
\usepackage{caption}

\usepackage{amsmath} 

\usepackage{siunitx}

\title[\cxo\ observations of \so]{Extended X-ray emission from the $z=4.26$ radio galaxy \so}

\author[Napier et al.]
{Kate Napier$^1$,
Adi Foord$^1$,
Elena~Gallo$^1$,
Gabriele~Ghisellini$^2$,
Edmund~Hodges-Kluck$^{3}$,\newauthor
Jianfeng~Wu$^4$,
Francesco~Haardt$^{5,6}$,
Benedetta~Ciardi$^7$ \\\\
  $^1$ Department of Astronomy, University of Michigan, 1085 S University Ave, Ann Arbor, MI 48109, USA \\
  $^2$ INAF, Osservatorio Astronomico di Brera, Via Bianchi 46, I-23807 Merate, Italy\\
  $^3$ NASA GSFC, Code 662, Greenbelt, MD 20771, USA\\
  $^4$ Department of Astronomy, Xiamen University, Xiamen, Fujian 361005, China\\
  $^5$ Universit\'a degli Studi dell'Insubria, Via Valleggio 11, I-22100 Como, Italy; INFN, Sezione di Milano-Bicocca, Piazza della Scienza 3, I-20126 Milano, Italy \\
  $^6$ INFN, Sezione di Milano-Bicocca, Piazza delle Scienze 3, 20123 Milano, Italy \\
  $^7$ Max Planck Institute for Astrophysics, Karl-Schwarzschild-Strasse 1, D-85748 Garching bei M{\"u}nchen, Germany\\
  }

\begin{document}
\maketitle
\begin{abstract}
We report on deep \textit{Chandra X-ray Telescope} imaging observations of \so, one of the few known radio galaxies at $z>3.5$. The X-ray counterpart is resolved into a core plus two off-nuclear sources that (combined) account for close to 30\%\ of the total X-ray flux. Their morphology and orientation are consistent with a diffuse, lobe-like nature, albeit compact hotspots cannot be ruled out. The broadband spectral energy distribution of \so\ can be reproduced with a jet model where the majority of the radio flux can be ascribed to synchrotron emission from the hotspots, whereas the (non-nuclear) X-ray emission is produced via Inverse Compton (IC) off of Cosmic Microwave Background (CMB) photons within the extended lobes. This scenario is broadly consistent with the expectation from highly magnetized lobes in a hotter CMB, and supports the view that IC/CMB may quench less extreme radio lobes at high redshifts.
\end{abstract}

\begin{keywords}
  galaxies: active --- galaxies: high-redshift --- galaxies: jets ---
  radiation mechanisms: non--thermal --- \hbox{X-rays}: galaxies
\end{keywords}

\section{Introduction}

Radio emission from jetted Active Galactic Nuclei (AGN) arises from the sum of the collimated, relativistic jets, plus the extended components, i.e. the hotspots and lobes. 
If the observer's line of sight lies within an angle $\sim$ $1/\Gamma$ of the jet axis, where $\Gamma$ is the jet bulk Lorentz factor, then the beamed emission from the jet dominates the spectrum, and the system is classified as a blazar. For larger viewing angles, isotropic extended emission from the lobes takes over, and we observe a powerful radio galaxy. 
Since, for each observed \bla, there must exist $\simeq 2\Gamma^2$ mis-aligned sources, the very existence of a dozen high-redshift ($z\gsim$4) \bla s \citep{yuan06,sbarrato15} indicates 
that a much larger population of mis-aligned, powerful, jetted AGN was already in place when 
the Universe was $\simlt$1.5 Gyr old.  

In spite of being well within the FIRST sensitivity threshold, this ``parent population'' remains elusive. More specifically, the expected number density of radio-loud quasars (as inferred from the \textit{Swift}-Burst Alert Telescope sample of luminous, massive \bla s \citealt{ajello09}), overestimates the number of observed luminous, radio-loud quasars (as identified by cross-matching the FIRST and Sloan Digital Sky Survey (SDSS) catalogs), by a factor $\sim$3 in the redshift bin $z=3-4$ and by a factor $>$10 between $z=4-5$ \citep{volonteri11}. Qualitatively similar conclusions are reached by \citet{kratzer15} (see also \citealt{haiman04} and \citealt{mcgreer09}). 
Possible reasons for this discrepancy include (i) heavy optical obscuration of high-$z$ radio galaxies, which would be missed by SDSS; (ii) a substantial drop in the average $\Gamma$ for high-$z$ sources, or; (iii) substantial, intrinsic dimming of the extended radio lobes at $z\simgt$3. 
The first scenario implies the existence of a large population of infrared-luminous, radio-loud quasars with no optical counterparts, while the second appears at odds with the notion of the most powerful jets being associated with high accretion rates, likely common in the early Universe. With respect to the third hypothesis, the idea has long been entertained of Cosmic Microwave Background (CMB) photons affecting the behavior of jetted AGN (e.g. \citealt{celotti+fabian04}). 

\citet{ghisellini14} explore specifically how the interaction between the CMB radiation and electrons within jet-powered hotspots and {lobes} affects the appearance of jetted AGN at different redshifts. The major result can be summarized as follows: Owing to its $(1+z)^4$ dependence, the CMB energy density starts to dominate over magnetic energy density within the lobes above $z$ $\simeq$3, thereby suppressing the synchrotron radio flux at higher $z$. At the same time, high-energy electrons will cool effectively by inverse Compton (IC) scattering off of CMB photons.
{Combined, these two effects result in a significant suppression of the radio lobes, accompanied by enhancement of the X-ray lobes, in high-$z$ quasars, which would then be {\it misclassified} as radio-quiet.}\\

From an observational standpoint, it is known that the radio lobes of low-$z$ radio galaxies can be associated with extremely low-surface brightness X-ray emission, likely due to CMB scattering (e.g. \citealt{hardcastle04, croston_hardcastle_05}). {If CMB-quenching of \hz\ radio lobes is indeed effective, we would expect the X-ray-to-radio flux ratio in lobes to increase with $z$, eventually leading to the disappearance of most radio-lobes.}
Direct confirmation of CMB quenching -- i.e. through detecting X-ray lobes around high-$z$ blazars and their parent population -- {faces major, inherent observational challenges}. For blazars and radio-loud quasars the core/jet emission is guaranteed to obliterate any diffuse GHz and X-ray emission (at any $z$; see e.g. \citealt{yuan03}; \citealt{siemiginowska03}; \citealt{lopez06}). At the same time, \hz\ radio-quiet quasars (the natural target of any such test) are highly risky targets, in that one has no a priori knowledge of which exactly are {misclassified as} -- as opposed to intrinsically -- radio-quiet.  

Based on the lack of spectral softening below 1 GHz (rest-frame), \citet{ghisellini15} conclude that {none} of the 13 known $z>$ 4 blazars show evidence for extended radio emission based on their GHz spectra, lending indirect support to the CMB-induced radio-quenching of high-$z$ AGN. 
Perhaps counter-intuitively, a direct, powerful test for this model lies with high spatial resolution X-ray observations of those exceedingly rare $z$ $>$ 3.5 radio galaxies \citep{debreuck10}. In the context of CMB quenching, detectable radio lobe emission in spite of the strong IC coolant provided by the CMB at $z\simgt$4 can be explained if the lobes of these systems are characterized by high compactness and/or high magnetic field, leading to an unusually strong enhancement of the synchrotron radio emission. For these systems, the CMB quenching model makes strong, testable predictions; that {(i) the radio lobes must also be X-ray sources; (ii) the X-ray lobes must be more luminous than for a comparably young radio galaxy at low redshift.} Among the few known \hz\ radio galaxies, \so\ ($z$=4.261; \citealt{spinrad95}) is the only one to have an associated, statistically significant (5$\sigma$) X-ray counterpart at $z>4$ (see \S\ref{sec:xmm} and Figure \ref{fig:map}).

In this Paper, we report on deep observations of \so\ with the \textit{Chandra X-ray Observatory}, aimed at testing the CMB-quenching model by resolving its X-ray emission on sub-arcsec scales. Archival radio and X-ray observations of the field of view of \so\ are presented and discussed in \S\ref{sec:xmm}; \S\ref{sec:cxo} describes the \cxo\ data reduction and analysis. An interpretation of the results in the context of the CMB quenching model is presented in \S\ref{sec:model}. We conclude in \S\ref{sec:discussion} with a discussion of our results vis-\`{a}-vis a broad assessment of the viability of IC/CMB quenching of \hz\ radio galaxies.

\section{Archival observations of \so\ in the context of the CMB-quenching model}\label{sec:xmm}

\begin{figure}
\centering{
\includegraphics[width=0.95\linewidth]{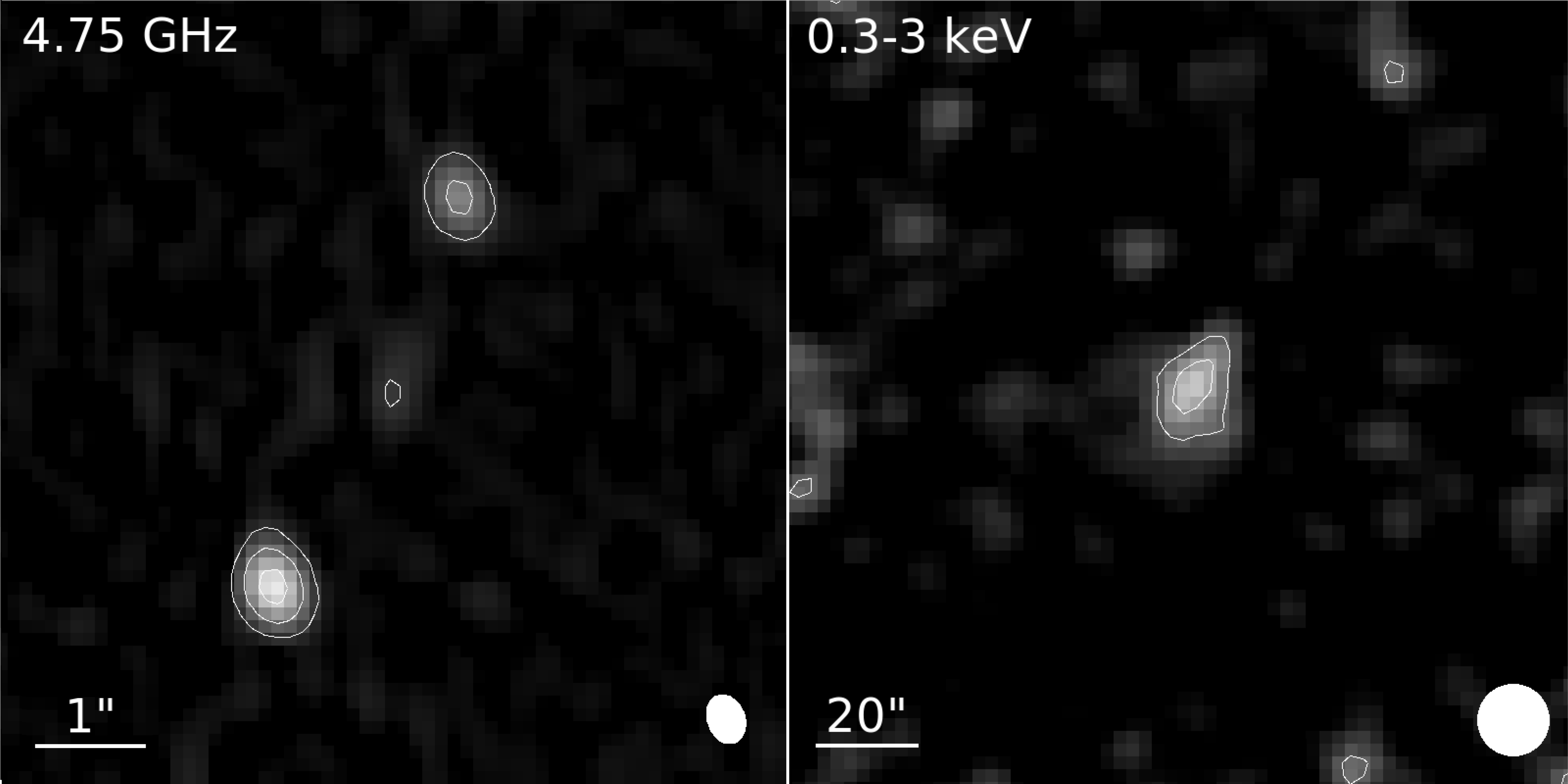}
\caption{Left: 5~GHz VLA image of \so. The filled white ellipse denotes the 0.4\arcsec\ beamsize and the contours are at 2, 10, and 30 mJy beam$^{-1}$. The rms noise is 0.11 mJy beam$^{-1}$. Right: 0.3-3~keV \textit{XMM-Newton} image (combined MOS and pn). The filled white circle represents the pn point-spread function: at this resolution, the X-ray image is consistent with a point source. The contours represent 3 and 6$\sigma$ above the background.}
\label{fig:map}}
\end{figure}

\begin{figure}
\centering
\includegraphics[width=\linewidth]{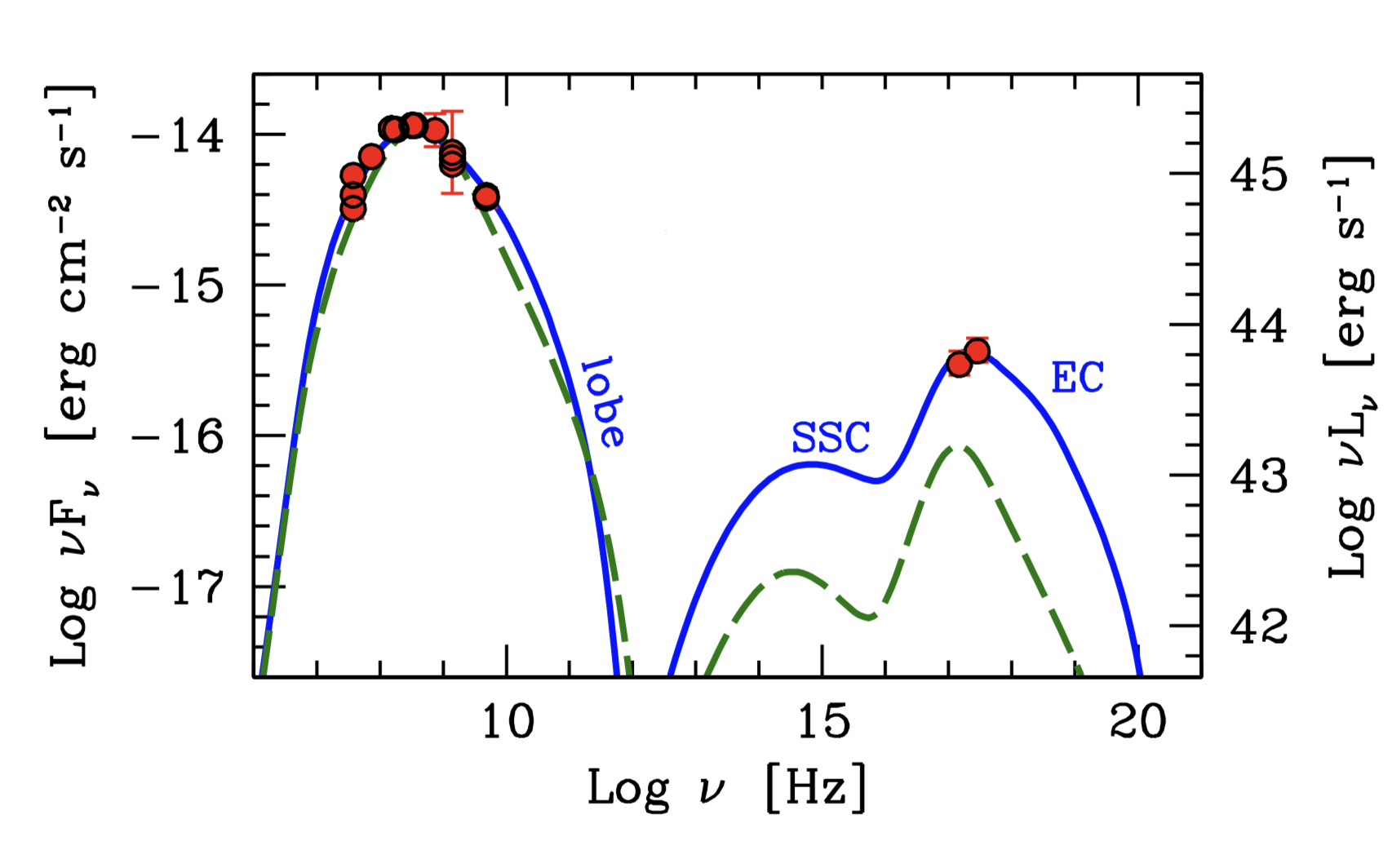}
\caption{Combined radio and X-ray SED of the $z$ = 4.261 radio galaxy \so, where the radio data points refer to the off-nuclear sources and the X-ray data points are based on the \xmm\ detection shown in the right panel of Figure \ref{fig:map}. The coloured curves represent two realizations of a {model} which attempts to reproduce the SED with emission from jet-powered {lobes}, following \citet{ghisellini09}. Highlighted are the Synchrotron Self-Compton (SSC) and External Compton (EC) components, where the latter include Inverse Compton scattering of the relativistic electrons off of (primarily) CMB photons (IC/CMB). The solid blue line represents a version of the model where 100\%\ of the X-ray flux can be ascribed to the lobes. By comparison, the dashed green line represents a model where only 25\%\ of the measured X-ray flux is ascribed to the lobes. Suppressing SSC and IC/CMB cooling to this extent requires a higher degree of magnetization within the lobes compared to the former case. }\label{fig:sed}
\end{figure}

The left panel of Figure \ref{fig:map} shows a 5~GHz image of \so\ \citep[from the Very Large Array (VLA) project AC374;][]{lacy94}. The source is composed of a faint core plus two off-nuclear radio sources, which we shall (temporarily) refer to as SE and NW ``lobes'', each separated by 1.8-2\arcsec\ from the core (i.e.,  $\simeq$ 12-14 kpc at $z = 4.26$). The field of view of \so\ falls within an archival \xmm-\textit{Newton} observation (ObsID 0204400301), performed in 2004 February. We extracted, reprocessed, filtered and analyzed the data with SAS v14.0. An X-ray source is clearly detected at position consistent with \so\ in the combined MOS+pn image, shown in the right panel of Figure \ref{fig:map}. 
The projected separation of the 5~GHz radio lobes is approximately 4\arcsec, implying that any co-spatial X-ray emission could not be resolved by \xmm-pn, whose Point Spread Function (PSF) Full Width Half Maximum (FWHM) is $<12.5$\arcsec. 
%
The two-point \xmm\ spectrum of \so\ (modeled with a redshifted, absorbed power law) is shown in Figure \ref{fig:sed}, together with the broadband radio spectrum of the extended (i.e., non-core) emission (based on publicly available aggregated data from \url{https://tools.ssdc.asi.it/}). The solid blue line represents a trial model run where the radio$+$X-ray Spectral Energy Distribution (SED) is modeled in terms of synchrotron, synchrotron self-Compton (SSC) and Inverse Compton (IC) radiation due to interaction of the lobes' electrons with photons from the disc, the external torus and the CMB (collectively referred to as External Compton, or EC), following \citet{ghisellini09}. This plot illustrates how emission from the \textit{lobes alone} could account for the entire X-ray and radio emission from \so, provided that the lobes are mildly magnetically dominated, i.e., have a magnetic-to-radiation energy density ratio $U_{\rm B}/U_{\rm rad}\simeq $4. 
Simply put, for radio synchrotron emission not to be quenched by IC scattering of CMB photons at this redshift, the synchrotron cooling process must be locally enhanced by supra-equipartition values of the magnetic field strength (unless otherwise noted, with  equipartition we refer to the balance between magnetic and \textit{radiation} energy density, rather than particle).
The corresponding, inferred total energy in magnetic field and particles within the lobes (about $2\times 10^{59}$ erg) would be typical of a very young system, whose lobes may still be expanding and are yet to reach equipartition (notice that knowledge of the size and radio flux of the lobes enables us to constrain the magnetic field value with no a priori assumption of equipartition in this system). While this is an extreme scenario, where virtually 100\%~of the X-rays detected by \xmm\ would arise from Comptonization of CMB photons within the (unresolved) lobes, 
this kind of exercise enables us to anticipate what the detection of extended X-ray emission from \so\ could imply in terms of model constraints.
E.g., if 25\%~of the total X-ray flux were emitted from the lobes (this case corresponds to the dashed green line in Figure \ref{fig:sed}), this would translate into unusually high magnetization levels, with $U_{\rm B}/U_{\rm rad}\simeq$17 and a B-field of $\simeq$ 1 mG, which is on the high end even for GHz-peaked sources (which are thought to be extremely young and tend to have projected linear sizes of $\simlt$1 kpc). This would also imply a total energy of $10^{60}$ erg within the lobes, which is typical of the most powerful radio galaxies, having $\sim$100 kpc-scale lobes. \\

We wish to stress that these models only serve to illustrate the role of IC/CMB vs. synchrotron cooling for a highly idealized case where the radio$+$X-ray SED is produced by the source extended lobes.

The situation is arguably more complex in the case of \so; a careful analysis of all the archival VLA images available for this system suggests that the majority of the off-nuclear radio emission at 5 GHz (observed-frame) emanates from \textit{point-like hotspots, rather than truly diffuse lobes}. Comparing two 5~GHz images with different beam sizes -- 1.4\arcsec\ and 0.4\arcsec\ FWHM --  and assuming each detected radio source is point-like in both images, one recovers that 78\%~(93\%) of the flux in the 1.4\arcsec\ beam is contained in the 0.4\arcsec\ beam for the SE (NW) source. That is, a strong majority of the emission comes from a $\sim$spherical volume of $\simlt$ 1.2 kpc in radius. Hereafter, we refer to these compact sources, which account for (on average) 85\%~of the measured off-nuclear flux at 5 GHz, as hotspots.\\

It should be noted that band C (4--8 GHz) is not ideally suited for detecting extended radio lobes, the more so at \hz. This is because, owing to the spectral steeping of the electron populations (i.e., the particle synchrotron lifetime scales with the inverse of the particle energy), emission at such high rest-frame frequencies (5 GHz corresponds to a rest-frame frequency of 21 GHz for \so) will die off very fast in the lobes; additionally, truly extended emission could be resolved out at high rest-frame frequencies. At the same time, however, lower rest-frame frequencies, such as 5~GHz (i.e., 1.4~GHz observed) typically lack the desired resolution. Indeed, the only available image of \so\ at 1.4 GHz (from the VLA FIRST survey) has $\simgt$5\arcsec\ FWHM; this yields a 1.4-5 GHz spectral index of about $+$1.5, which confirms that the 21~GHz (rest-frame) emitting-plasma should not be expected to contribute substantially to any extended, lobe-like emission.

Mindful of these caveats, we turn to \cxo, and the sub-arcsec resolution of its Advanced CCD Imaging Spectrometer (ACIS), to constrain the nature of \so's X-ray counterpart on scales that are comparable to those of the observed radio hotspots. 
\begin{figure}
    \centering
   {\includegraphics[width=0.35\textwidth]{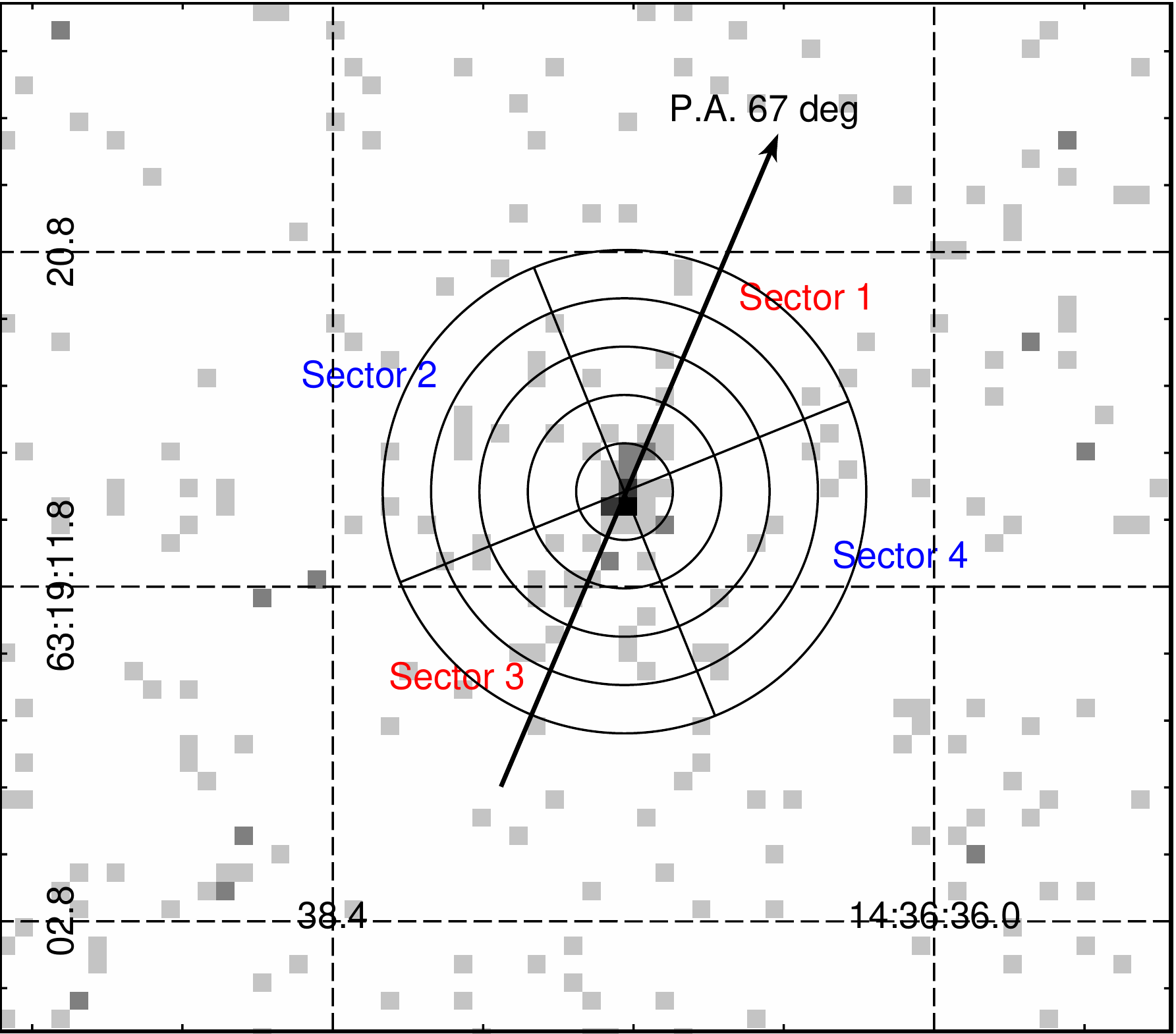}}
   {\includegraphics[width=0.425\textwidth]{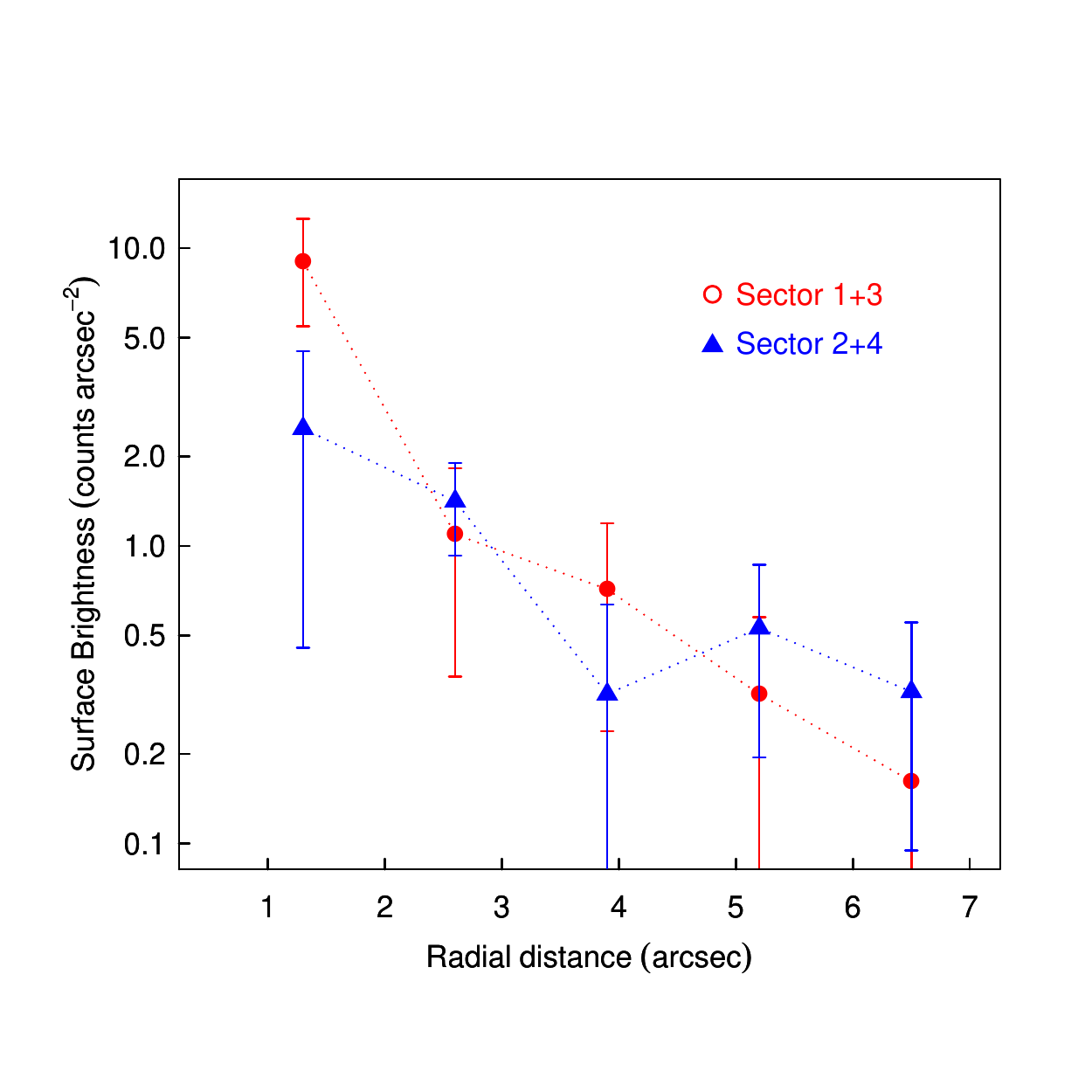}}
    \caption{{Top: \cxo\ image of the the field of view of \so, in the 0.5-8 keV band, obtained by combining three different visits, for a total of approximately 100 ks of net exposure. The raw image is overlaid with a series of five concentric annuli extending out to about 8 arcsec; these are divided into four, pie-shaped sectors: sector 1 and 3 are aligned with the position angle (P.A.) of the radio hotspots ($\simeq$67 degrees; see Figure 1, left panel); sector 2 and 4 are perpendicular to it. 
    Bottom: Comparison of the source surface brightness profiles as measured along (sector 1$+$3, in red) and across (sector 2$+$4, in blue) the P.A. of the radio knots. The former reveals an excess of counts within $\simlt$2 arcsec of the center, suggesting that \so\ displays extended X-ray emission in the same general direction as its radio counterpart. } \label{fig:data_psf_prof}}
\end{figure}
\section{Chandra data Reduction and Analysis}\label{sec:cxo}
We observed \so\ over three epochs (Observation IDs 20033, 19954, 18106; PI: Gallo) for a total of 100 ks with ACIS-S. The source was placed at the aim point of the back-illuminated S3 chip. The combined exposure was established by running {\sc marx} simulations so as to ensure that extended X-ray emission on arcsec-scales, if present, could be detected if they each contributed $\simgt$10\%~to the total flux measured by \xmm\ (see \S\ref{sec:ratio} for a quantitative analysis). 

We reduced the data using the \textit{Chandra} Interactive Analysis of Observations ({\sc ciao}), v4.10.  We first reprocessed the data and created new level 2 event and bad pixel files using the {\sc ciao} script \verb+chandra_repro+. The astrometry was corrected by cross-matching the centroid positions of all the point-like X-ray sources (outside of the main target positional error circle) detected by {\sc ciao}\textquotesingle s \verb+wavdetect+ to the SDSS Data Release 8 catalog. For each of the three observations, we ran \verb+wavdetect+ on the reprocessed level 2 event file and required the source match radius to be 2\arcsec\ or fewer. Each image required an astrometric shift of no more than 0.4\arcsec.  Finally, we re-projected the three observations to the same tangent point and merged them. 

{\sc ciao}'s \verb+wavdetect+ clearly detects an X-ray source within the nominal target error circle. A total of $52\pm 8$ net counts were estimated using a 5\arcsec\ radius circular region for the source, and a (source-free) 20-30\arcsec\ annulus for the background, both centered at the \verb+wavdetect+ source centroid position.  
Using {\sc xspec} version 12.10.1 \citep{Arnaud1996}, we fit \so's 0.5-8.0 keV spectrum using an absorbed redshifted power-law model (\verb+phabs+*\verb+zphabs+*\verb+zpow+). We fixed the hydrogen-equivalent Galactic column (i.e., the photoelectric absorption component {\tt phabs}) to $N_{\rm H} = 1.4\times10^{20}$~cm$^{-2}$ (as given by the {\sc~colden} tool for the target's coordinates; see also \citealt{Kalberla2005}) and the redshift to $z = 4.26$ and implemented the Cash statistic ({\tt cstat}; \citealt{Cash1979, Brightman2012}) in order to best assess the quality of our model fits. 
The spectrum is best fit by a power law index $\Gamma = 1.51 \pm 0.3$ and a negligible intrinsic hydrogen-equivalent column ($N_{\rm H}<10^{20}$ cm$^{-2}$). The corresponding, measured 0.5-8.0 keV flux is $(5.8^{+0.33}_{-3.28}) \times10^{-15}$~erg~cm$^{-2}$~s$^{-1}$, consistent, within the errors, with \xmm's (i.e., $3.8\pm 1.2 \times 10^{-15}$ erg sec$^{-1}$ cm$^{-2}$). 
\subsection{X-ray source morphology}\label{sec:morph}
To assess the key question whether \so\ is associated with a point-like or extended X-ray source, we implemented three methods, with increasing levels of refinement.  We first used \verb+srcextent+, a {\sc ciao} tool that estimates the extent of a source image and, if a PSF image is supplied, returns whether or not the source is extended within a given confidence interval.  We simulated the source PSF (at 1.5 keV) using the Chandra Ray Tracer (ChaRT) software to generate the ray files and {\sc marx}  \citep{Davis2012} to project the rays onto the detector plane (Figure \ref{fig:data_psf_prof}).  We ran \verb+srcextent+ several times, each with different choices for the source shape and the region file encompassing the source. 
Although \verb+srcextent+ determined that \so's X-ray counterpart is not extended (with 90$\%$ confidence) for the majority of our trials, its final assessment depended sensitively on the chosen source extraction region (circle vs. ellipse) and shape (Gaussian vs. disc), and was thus deemed inconclusive for our purposes.

Second, we measured the source surface brightness profiles {along and across the position angle (P.A.) defined by the direction of the known radio hotspots, {i.e. $\simeq$67 degrees (measured counterclockwise relative to the axis in the direction of decreasing R.A.)}, as described in Figure \ref{fig:data_psf_prof}. This analysis suggests that \so\ displays excess X-ray emission along the same direction as the radio hotspots.}
  
To characterize the nature of this excess emission, we analyzed the observations with the {\sc python} tool \BAYMAX{} (\textbf{B}ayesian \textbf{A}nal\textbf{Y}sis of \textbf{M}ultiple \textbf{A}GNs in \textbf{X}-rays; \citealt{foord19}). \BAYMAX{} is capable of ascertaining whether a given \cxo-detected source is better described by a model composed of one or multiple point sources by calculating the Bayes Factor ($BF$), which represents the ratio of the likelihood of the observed data, given two different models. In brief, \BAYMAX{} takes calibrated events from reprocessed \emph{Chandra} observations and compares them to simulations based on single and multiple point source models. The properties of the \emph{Chandra} PSF are characterized by simulating the PSF of the optics with {\sc marx}. For a source with multiple observations, \BAYMAX{} first models the PSF of each observation and calculates the likelihoods for each observation individually (which is expected to depend on the detector position and start time of the observation), and then fits for astrometric shifts between different observations of the same source.  The adopted PSF model is energy-dependent, such that it accounts for the spread at higher energies. Below, we review the details when running \BAYMAX{} on observations of \so, however for a more detailed review on the statistical techniques behind \BAYMAX{}'s calculations, we refer the reader to \citealt{foord19}. 

For \so, we used \BAYMAX{} to compare a single point source model ($M_{\mathrm{S}}$) to a triple point source model ($M_{\mathrm{T}}$). For both models, each photon is assumed to originate from either a point source component or the background. Thus, $M_{\mathrm{S}}$ and $M_{\mathrm{T}}$ are parameterized by vectors $\theta_{S}=[\mu, \log{f_{BG}}, \Delta x_{k}, \Delta y_{k}$] and $\theta_{T}=[\mu_{{N}}, \log{f_{n}}, \log{f_{BG}}, \Delta x_{k}, \Delta y_{k}]$. Here, $\mu_{N}$ represents the location for $N$ point sources ($N=[1, \dots,  N]$); $f_{BG}$ represents the ratio of counts between the background and the combined counts from all point source components; $f_{n}$ represents the ratio of the total counts between a given point source and the primary point source ($n=[1, \dots, N-1]$); and $\Delta x_{k}$ and $\Delta y_{k}$ account for the translational components of the relative astrometric registration for the $k=[1, ..., K-1]$ observation (where $K=3$ for \so). 
For a single point source, the probability that a photon observed at sky coordinate ($x$,$y$) with energy $E$ is described by the PSF centered at $\mu$ is $P(x,y \mid \mu, E)$, while for a triple point source the total probability is $P(x, y \mid \mu_{N}, E, f_{n}, f_{\mathrm{BG}})$.  Regarding the background, we assume that photons are uniformly distributed across a given region, such that the probability that a photon observed at location $x$,$y$ on the sky with energy $E$ is associated with a background component is $P(x,y \mid f_{BG},E)$. Here, $f_{BG}$ represents the ratio of counts between the background and the combined counts from all point source components. 
\begin{figure}
    \centering
    \includegraphics[width=8cm]{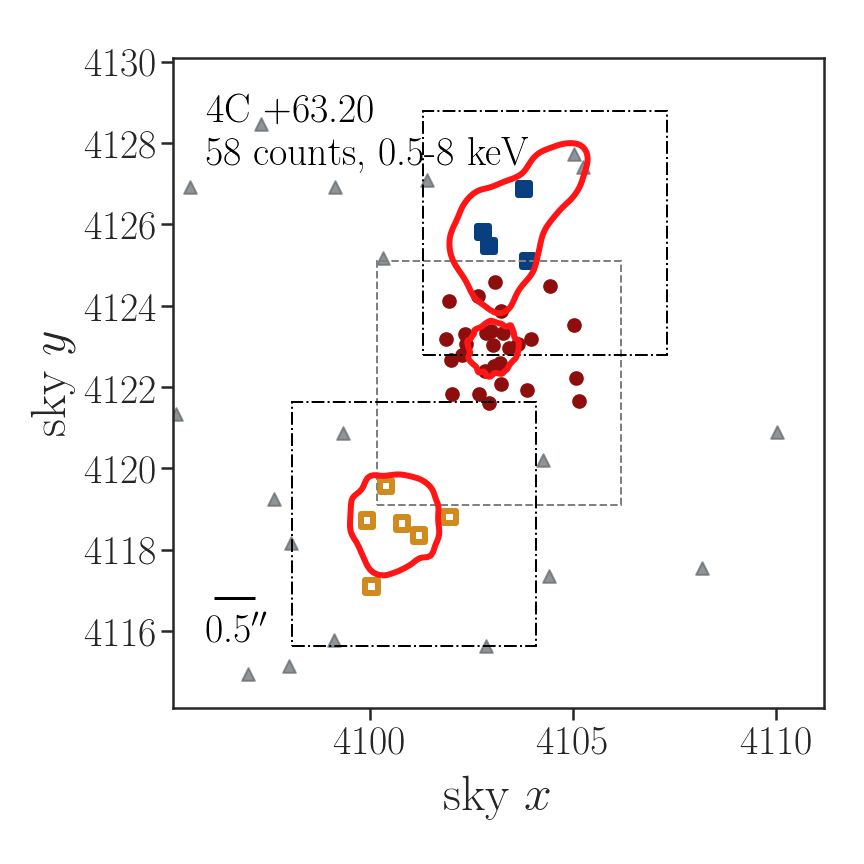}
    \caption{The combined 0.5-8 keV data set for \so, as seen by \BAYMAX{} for a triple point source model ($M_{\mathrm{T}}$). Each count is probabilistically assigned to a specific model component. In the case of informative priors, each component is limited by the known location of the radio core and off-nuclear sources, such that each count is allowed to vary within the dashed and dash-dot boxes, respectively. When using non-informative priors, component counts were allowed to be anywhere within the image field-of-view. The resulting 68\%\ confidence intervals for the best-fit sky $x$ and sky $y$ positions for the core and the off-nuclear components are illustrated by the solid red contours. Counts most likely associated with the core are denoted by dark red dots, while counts most likely associated with the NW and SE radio sources are denoted by blue (filled) and orange (open-faced) squares. Counts most likely associated with background are shown as faded grey triangles.\label{fig:bestfit_BAYMAX}}
\end{figure}%

All prior distributions of $\mu$ for both $M_{S}$ and $M_{T}$ are described by continuous uniform distributions. When using non-informative priors, the coordinates of each $\mu$ are allowed to be anywhere within a given region centered on the X-ray source centroid position; when using informative priors, the coordinates of each $\mu$ are defined by the locations of the core and off-nuclear radio sources as determined in the archival VLA radio map shown in Figure \ref{fig:map}. We note that our informative prior distributions for $\mu$ are wide enough to account for the relative astrometric shifts between the \emph{Chandra} and VLA observations ($>1\arcsec$). The prior distribution of $\log{f_{BG}}$ is described by a Gaussian distribution, $N$($\mu_{BG}$, $\sigma^{2}_{BG}$), where $\mu_{BG}$ is estimated by evaluating the count-rate in 10 random and source-free regions within a 20\arcsec\ radius of the X-ray source centroid position.  We set $\sigma_{BG}$ to 0.5, allowing for \BAYMAX{} to easily move in parameter space.  For $M_{T}$, the prior distributions of $\log{f_{n}}$ are described by uniform distributions and are constrained between $-2$ and $0$, accounting for a large range of possible count fractions between the off-nuclear sources and core. Lastly, the prior distributions of $\Delta x_{k}$ and $\Delta y_{k}$ are described by a uniform distribution constrained between $\delta \mu_{\mathrm{obs}}-3$ and $\delta \mu_{\mathrm{obs}}+3$, where $\delta \mu$ represents the difference between the observed central X-ray coordinates of two given \cxo\ observations. Here, we define $\Delta x_{k}$ and $\Delta y_{k}$ relative to the longest observation (ObsID: 18106).
We analyzed the photons contained within a $16\times 16$ sky-pixel region that is centered on the centroid position (corresponding to $>$95\%~of the encircled energy radius for the $0.5$-$8$ keV photons, and encompassing the majority of the X-ray emission expected to be associated with core and off nuclear sources; Figure \ref{fig:bestfit_BAYMAX}).  We masked the known asymmetric \emph{Chandra} PSF feature in each observation (which sits approximately 0\farcs7 from the center of the core; \citealt{Juda&Karovska2010}), as our PSF model does not take this asymmetry into account. We found $BF$ values {strongly in favor of the triple point source} when using both non-informative ($\ln{BF} = 2.57\pm 0.60$) and informative priors ($\ln{BF} = 4.85\pm0.64$).  Error bars are defined by the spread of $BF$ values after running \BAYMAX{} 100 times, while $BF$ are classified as ``strong'' via false positive tests (see \citealt{foord19}). 
%
\begin{figure}
    \includegraphics[width=1.02\linewidth]{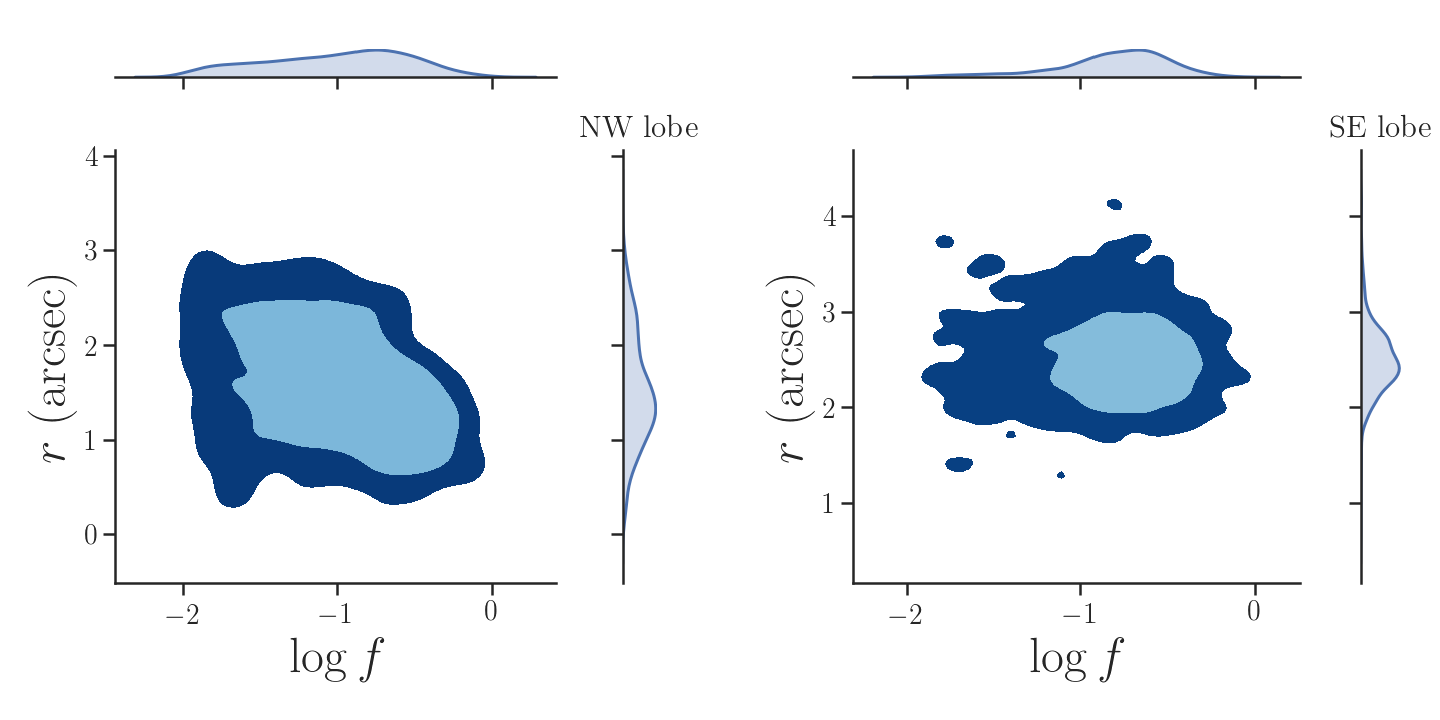}
    \includegraphics[width=0.5\linewidth]{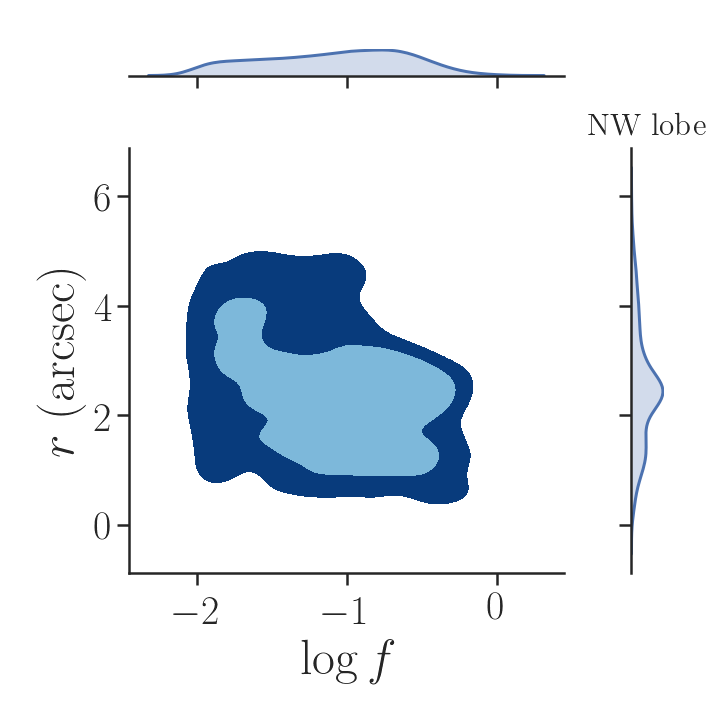}\includegraphics[width=0.5\linewidth]{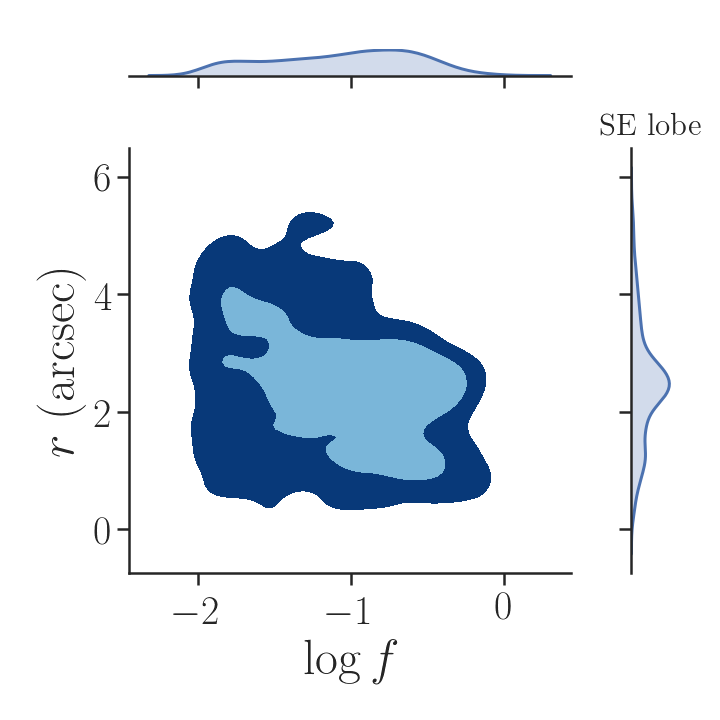}
    \caption{Joint posterior distribution for the inferred separation $r$ (in arcsec) and the count ratio (in units of $\log{f}$) for the NW (left panel) and SW (right panel) X-ray source, as determined by \BAYMAX{} for the case of informative priors (top panels) and non-informative priors (bottom). The 68 and 95\%\ confidence intervals are shown in light and dark blue contours, respectively, while the marginal distributions are shown along the border. The NW and SE X-ray sources are estimated to contribute $\sim$11 and $\sim$17 percent of the total counts, respectively.\label{fig:jointposteriors_BAYMAX}}
\end{figure}%
\begin{figure*}
    \centering
    \includegraphics[width=\textwidth]{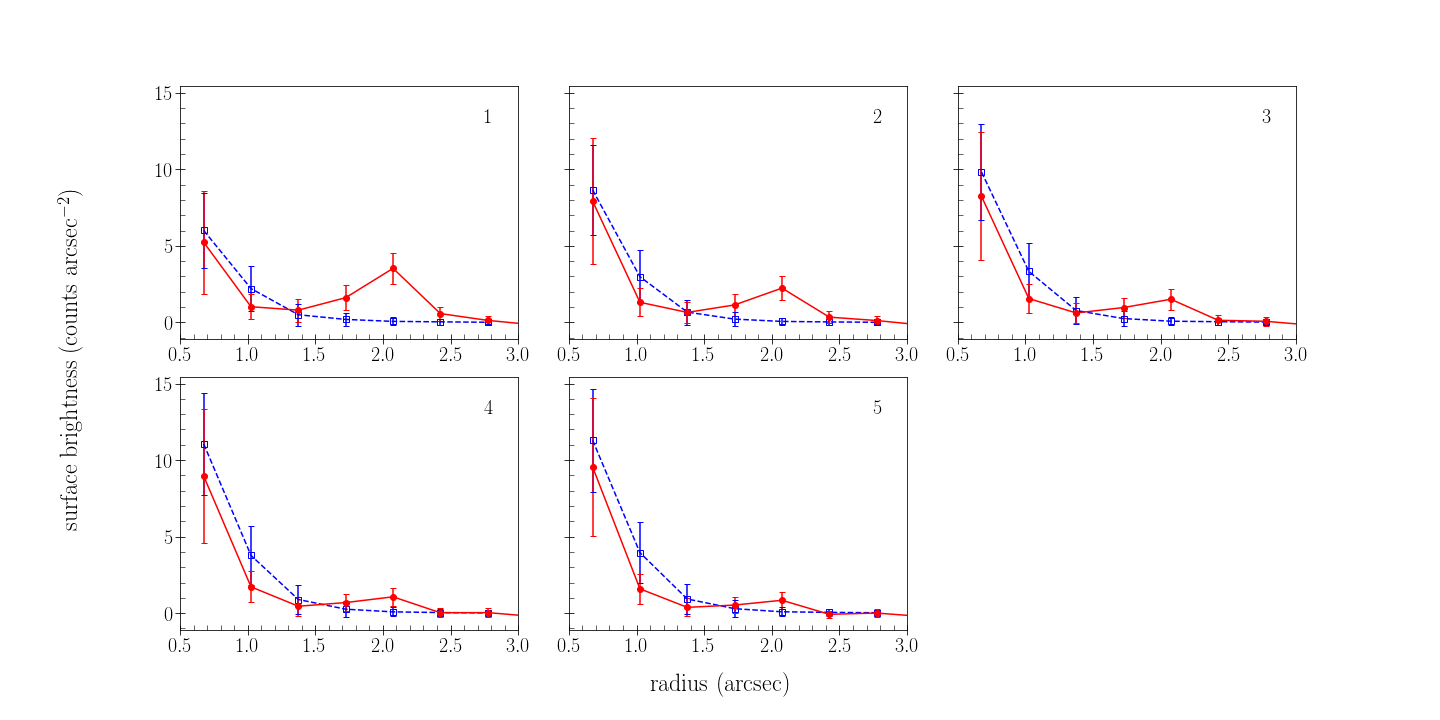}
    \caption{{Comparative radial surface brightness profile analysis, where the profile of a single PSF (shown in blue) is compared to five sets of simulated data profiles (shown in red). Each data set is composed of three simulated PSFs with centroid positions fixed to \so's core, SE and NW radio sources, each having a different ratio of the core ($X$) to off-nuclear sources X-ray counts ($Y=(100 - X)/2$, with $Y=25, 15, 10, 5, 3$ corresponding to panel 1, 2, 3, 4 and 5). Based on a statistical analysis (see main text), we conclude that, collectively, the SE and NW X-ray sources must contribute to $\simgt 10$ percent of the total measured X-ray flux from \so\ in order to be significantly recovered in a 100 ks \cxo\ exposure.} \label{fig:sim_rad_prof}}
\end{figure*}%
As shown in Figure~\ref{fig:jointposteriors_BAYMAX}, the best-fit coordinates for the SE component using non-informative priors are consistent with those found using informative priors (albeit, with larger relative uncertainties). Although this remains true for the best-fit coordinates of the NW component, the coordinates are not as well-constrained when using non-informative priors. With informative priors, we find the NW and SE components separated from the core with $r=1.59\arcsec_{-0.68}^{+0.69}$ and $r=2.43\arcsec_{-0.37}^{+0.37}$, and count ratios of $\log{f}=-0.90_{-0.60}^{+0.40}$ and $\log{f}=-0.73_{-0.39}^{+0.20}$, respectively.

{Posterior distributions for the P.A. of the X-ray lobes were generated starting from the posteriors for the best-fit sky coordinates of each lobe. Both informative and non-informative priors yield P.A. posterior distributions that are consistent with the P.A. of the radio lobes ($\simeq 67$ degrees), albeit non-informative priors yields a much broader distribution. Specifically, the median and 68\% confidence interval values of the off-nuclear X-ray source P.A. are $67.40\pm 8.75$ ($61.89\pm 13.92$) and 8.40 ($35.48$) degrees for the informative (non-informative) priors, respectively.}\\

Based on our three assessments, \textit{we conclude that \so\ is associated with an extended X-ray source, with a SE to NW orientation consistent with the orientation of the known radio hotspots}. The best-fit value for the fractional counts of the NW and SE components (defined as the median of the posterior distributions) are $\sim$11 and 17 percent, respectively. 
As to whether the SE and NW X-ray sources are point-like, and thus consistent with a hotspot vs. extended/lobe-like nature, the posterior distribution for the separation allows for both interpretations (as well as a combination of the two).
\subsection{Nuclear-to-extended flux ratio}\label{sec:ratio}
We quantify the fractional contribution of \so's off-nuclear components to the total measured X-ray flux with two methods: first, we compare a suite of simulated radial surface brightness profiles to the profile of a single PSF; second, we turn again to a Bayesian methodology.
As a first pass, we use {\sc marx} to simulate \so\ as the sum of three PSFs -- for the core, NW and SE component -- with varying degrees of flux contribution assigned to the off-nuclear components, i.e.: 25, 15, 10, 5 and 3 per cent (we do not explore the case where each source contributes different percentages). The nominal positions of the off-nuclear sources were chosen to match the measured peak of the SE and NW radio sources at 5 GHz (i.e., the hotspots). Given the low number of counts, the spectral shape of each simulated PSF was set to the best-fit model for the entire X-ray source. For each choice of $Y$ (with $Y=25, 15, 10,5, 3$ percent), we simulated three PSFs 100 times each, concatenated them and converted them to event files, using {\sc marx}\textquotesingle s \verb+marx2fits+, and performed a pixel-by-pixel averaging of the 100 event files to obtain a single, representative event file for each choice of $Y$. The appropriate blank-sky background files were generated, combined and aligned with the simulated source events file using the {\sc ciao} script \verb+blanksky+ and \verb+blanksky_sample+; {\sc marx}\textquotesingle s \verb+marxasp+ was used to create an aspect solution file for the internal dither model.   

{We carried out a comparative radial surface brightness analysis between the simulated data set profiles and the profile of a single PSF (shown in Figure \ref{fig:sim_rad_prof}). We fit various functional shapes to the single PSF's profile and found that the residuals were minimized for an exponentially declining function.  Then we ran a $\chi$-square test between this exponential fit and the radial profiles of the five simulated data sets, where a p-value of $0.05$ was chosen as the threshold for a statistically significant difference between the profiles. Based on this analysis, we confirm that any extended emission that contributes to less than 10 percent (i.e., panel 4 and 5 in Figure \ref{fig:sim_rad_prof}) of the total flux from \so\ could not be significantly detected in our merged, 100 ks \cxo\ observation (this, however, assumes that the two extended sources, each located at a position consistent with the radio hotspots, contribute equally to the extended flux).}\\

\begin{figure}
    \centering
    \includegraphics[width=\linewidth]{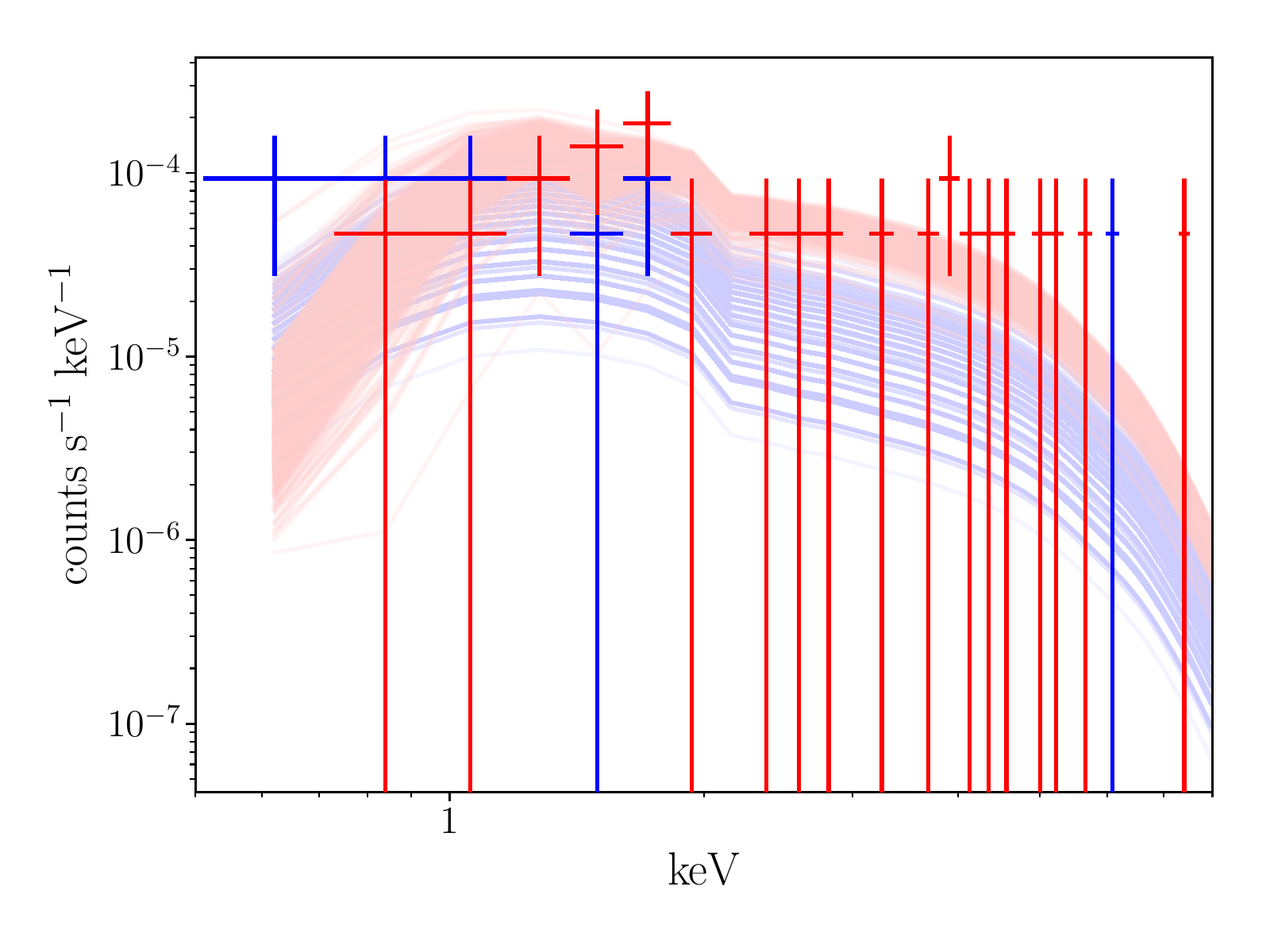}
    \caption{Fits to the 1,000 \BAYMAX{}-generated spectra of the core (shown as red solid lines) and the SE and NW X-ray sources (shown as blue solid lines). Fits are performed using a redshifted, absorbed power-law model with fixed slopes ($\Gamma=1.5$ and 1.8 for the SE/NW sources and core, respectively).  Over-plotted to the fits is one (out of 1,000) simulated spectrum for the core (red points and error bars) and one for the off-nuclear sources (blue points and error bars). \label{fig:fits} }
    \label{fig:spec_fit}
\end{figure}

To estimate the actual flux from each X-ray source, we used \BAYMAX{} to create 1,000 spectral realizations for each component within $M_{\mathrm{T}}$ (i.e, the 3-component model), plus the background.  This is done by probabilistically assigning each count to a specific model component, where the probabilities drawn from $\theta_{\mathrm{T}}$. 
Spectral fits are carried out within XSPEC and evaluated with the Cash statistic. We model both the spectral realizations of the core and the combined off-nuclear emission as an absorbed power-law ({\tt phabs}$\times${\tt zphabs}$\times${\tt pow}), where we fix the Galactic hydrogen-equivalent column density to $1.4\times10^{20}$~cm$^{-2}$ and the redshift to $z = 4.261$. We also fix the intrinsic absorption parameter ({\tt zphabs}) to the Galactic value when fitting the SE and NW sources.
Unsurprisingly, for both we find that the models where $\Gamma$ is fixed (to 1.8 and 1.5 for the core and the off-nuclear sources, i.e. consistent with typical AGN and Comptonization spectra, respectively) are statistically preferable to those where $\Gamma$ is free to vary (this is quantified by comparing the distributions of the best-fit $C_{\mathrm{stat}}$ values; see Figure \ref{fig:spec_fit}). 
For both the core and the off-nuclear sources, we created distributions of spectral parameters based on the best-fit values from each fit. The X-ray flux and luminosity values are determined by the median of the distributions, where the quoted errors represent the 68\%~confidence levels. 
For the core, we infer an intrinsic absorption column of $N_{\mathrm{H}} =  3.6^{+1.2}_{-0.9} \times 10^{23}$ cm$^{-2}$ and a total observed $0.5$--$8$ keV flux of $3.7^{+0.3}_{-0.3} \times 10^{-15}$ erg s$^{-1}$ cm$^{-2}$. For the SE/NW sources, the observed $0.5$--$8$ keV flux is $9.8^{+5.3}_{-4.0} \times 10^{-16}$ erg s$^{-1}$ cm$^{-2}$ s$^{-1}$. We adopt the above values in the following Section, for the purpose of modelling \so's core and off-nuclear X-ray emission. 
\section{Spectral Energy Distribution modelling}\label{sec:model} 
The high redshift of the source ($z$ = 4.26) implies that the radiation energy density of the CMB is enhanced by a factor $(1+z)^4\simeq 760$ with respect to the local value. As a consequence, it is unavoidable that IC scattering off of CMB photons (IC/CMB) must play an enhanced role in cooling the system's relativistic electrons compared to an otherwise equal system at $z=0$. Here, we aim to quantify the contribution of this process in \so\ and, by extension, assess its efficacy as a viable mechanism to quench radio emission from the lobes of jetted AGN at high redshifts. 

The complete SED of \so\ is shown in Figure \ref{fig:model}. Data points (other than X-ray) were retrieved from \url{https://tools.ssdc.asi.it/}; to those, we add the core (red butterfly) and extended sources (orange diamond) spectrum/flux measured by \cxo/\BAYMAX. In line with previous work (e.g, \citealt{ghisellini09,mocz11, ghisellini14,wu17}), we model the system SED as the sum of the following components: (i) an accretion disc and an absorbing torus; (ii) a (misaligned) relativistic jet;  (iii) the combination of two 
compact hotspots (assumed at rest); (iv) the combination of two extended lobes. For practical purposes, however, we will only consider models where \textit{either} the hotspots or lobes contribute to the extended emission. We describe each model component in turn next. 
\subsection{Disc-torus emission}
The dotted black line in Figure \ref{fig:model} corresponds to the disc$+$torus model spectrum. The (rest-frame) optical-UV emission arises from a standard, geometrically thin, optically thick disc (higher energy peak); the disc is typically assumed to be surrounded by a dusty torus, which absorbs the disc's optical radiation and re-emits in the IR band as 300~K blackbody. The torus is assumed to have an opening angle (with respect to the normal to the disc) of $\sim 45^\circ$, and to reprocess nearly half of the disc flux. The line of sight towards the system is assumed to be well within the torus funnel, such that the observed disc emission is not significantly absorbed by it. 
Owing to the lack of far-IR data for this source, the torus emission (lower energy peak) is effectively unconstrained (this has a negligible impact on our conclusions, as the contribution of torus photons for the overall EC process is minimal). 

Attempting to reproduce the Wien's tail of the disc multi-colour blackbody (as the peak of the disc spectrum is not well sampled by the data) yields a luminosity $L_{\rm d}=1.5\times 10^{46}$ erg s$^{-1}$, and black hole mass $M=8\times 10^8 M_\odot$. 
%
\subsection{Jet emission}
The model we adopt is described in detail in \citet{ghisellini09}, with the only difference that the jet of \so\ is assumed to be slightly misaligned with respect to the line of sight, so that, unlike for blazars, it does not contribute appreciably to the detected radio emission.  
The model assumes that most of the (non--thermal) jet radiation arises from a single zone, where a fraction of the total jet power (kinetic$+$magnetic) is dissipated. 
The emitting zone is assumed to be spherical, with radius $R=\psi R_{\rm diss}$,
where $R_{\rm diss}$ is the distance from the black hole  
and $\psi=0.1$ rad is the semi-aperture angle of the conical jet. 
The emitting region moves with velocity $\beta c$, corresponding to a bulk Lorentz factor $\Gamma$.
Throughout the region, which is embedded in a tangled magnetic field of strength $B$,
relativistic electrons are injected at a rate $Q(\gamma)$, with total power $P^\prime_{\rm e}$
(as measured in the co-moving frame). The electron energy distribution $Q(\gamma)$ is taken as a smoothly connected double power-law, with 
slopes $s_1$ and $s_2$ below and above the break energy $\gamma_{\rm b}$, respectively: 
\begin{equation}
Q(\gamma)  \, = \, Q_0\, \frac{ (\gamma/\gamma_{\rm b})^{-s_1}}{1+
(\gamma/\gamma_{\rm b})^{-s_1+s_2} }\quad {\rm cm^{-3} s^{-1}}.
\label{eq:qgamma}
\end{equation}
This distribution extends between $\gamma_{\rm min}\simeq 1$ and $\gamma_{\max}$, where the exact value of $\gamma_{\rm max}$ is not critical since we always assume that $s_1\leq 1$, $s_2>2$.
The particle density distribution, $N(\gamma)$, is found by solving the continuity equation at a time equal to the emitting region light-crossing time, accounting for radiative losses through synchrotron and IC processes, as well as e$^\pm$ pair production. Seed photons for IC cooling are provided by the disc, the torus and the CMB (with the latter largely dominating). The jet carries power in the form of Poynting flux ($P_{\rm B}$) and
electron$+$proton bulk motion ($P_{\rm e}$ and $P_{\rm p}$). 
Additionally, we account for the power $P_{\rm r}$ that is spent by the jet
to produce the observed radiation. The jet is taken to be misaligned with respect to the line of sight by an angle $\theta_{\rm v}=10$ degrees; 10 percent of its total jet power is assumed to power the extended lobe components \citep{ghisellini14b}. {Albeit arbitrary, this choice can be thought of as a compromise between the source number statistics and intrinsic jet power, in the sense that the probability of seeing a source at lower inclination then 10 degree is very small (and would imply the existence of a much larger parent population); at the same time, larger angles would imply stronger de-beaming, and hence higher intrinsic jet power. }

The jet parameters are tuned to best reproduce the observed core X-ray emission. For reference, they are well within the average values inferred for a large sample of {\it Fermi}-detected flat spectrum radio quasars luminosities and redshifts comparable to \so's
(\citealt{ghisellini14b,ghisellini15}). A complete list of the adopted parameters is given in Table \ref{tab:jet}. 
\subsection{Extended emission: hotspots and lobes}
The \cxo\ observations presented here have established that $\sim$30\%~of the X-ray emission from \so\ can be ascribed to two off-nuclear components; nevertheless, the angular separation and low number of counts of both off-nuclear X-ray sources allow for both a hotspot and/or a lobe interpretation.  At the same time, as discussed in \S\ref{sec:xmm}, since the source is not resolved at low radio frequencies (e.g., $\simlt 1$ GHz, or $\simlt$ 5 GHz rest-frame) the intrinsic amount of truly diffuse, lobe radio emission cannot be estimated. Our best guess is based on the comparison of the available maps at 5 GHz, where the majority of the emission emanates from a pair of hotspots, with only about 15\%~likely arising from the diffuse lobes (to be clear, this does not imply the absence of extended radio lobes in \so, nor does it imply that any lobe emission only contributes to 15\%~of the total radio flux, as any extended lobe would hardly be detectable at such high rest-frame frequency, due to the short synchrotron life-time of the emitting particles). 

For the purpose of our modelling, thus, we consider two extreme (mutually exclusive) scenarios: I. a ``pure hotspot'' model, in which all of the (off-nuclear) radio $and$ X-ray emission arises from compact, 2.4 kpc-scale (1.2 kpc radius) hotspots, and, II. a ``pure lobe'' model, where all of the (off-nuclear) X-ray emission, along with 15\%~of the radio, arise from extended, 12 kpc-scale (6 kpc radius) lobes. For reference, $\simeq$2.4 kpc corresponds to the beam size of the highest resolution VLA image available (i.e., 0.4\arcsec), where the hotspots are only marginally resolved, whereas 12 kpc corresponds to the NW hotspot-radio core distance (since a spherical geometry approximation likely overestimates the lobe volume, we chose to adopt the smallest projected size for the radius; although the inferred model parameters are sensitive to this choice, the relative ratios, which we are mostly interested in, are not).

In both cases, the extended X-ray emission is produced via EC, that is, IC scattering off of disc, torus and (mainly) CMB photons. This mechanism competes with the standard synchrotron cooling that is responsible for generating the radio spectrum, as well as the SSC component at intermediate frequencies. Attempting to simultaneously reproduce the extended X-ray emission with either model yields an estimate of the local magnetic-to-CMB radiation energy density ($U_{\rm B}/U_{\rm rad}$), as well as the total energy in magnetic field ($E_{\rm B}$) and particles ($E_{\rm e}$). In turn, different degrees of magnetization correspond to different relative contributions of the SSC vs. EC component. We discuss the adopted parameters for each model in turn next. 
\begin{figure}
    \centering
    \includegraphics[width=\linewidth]{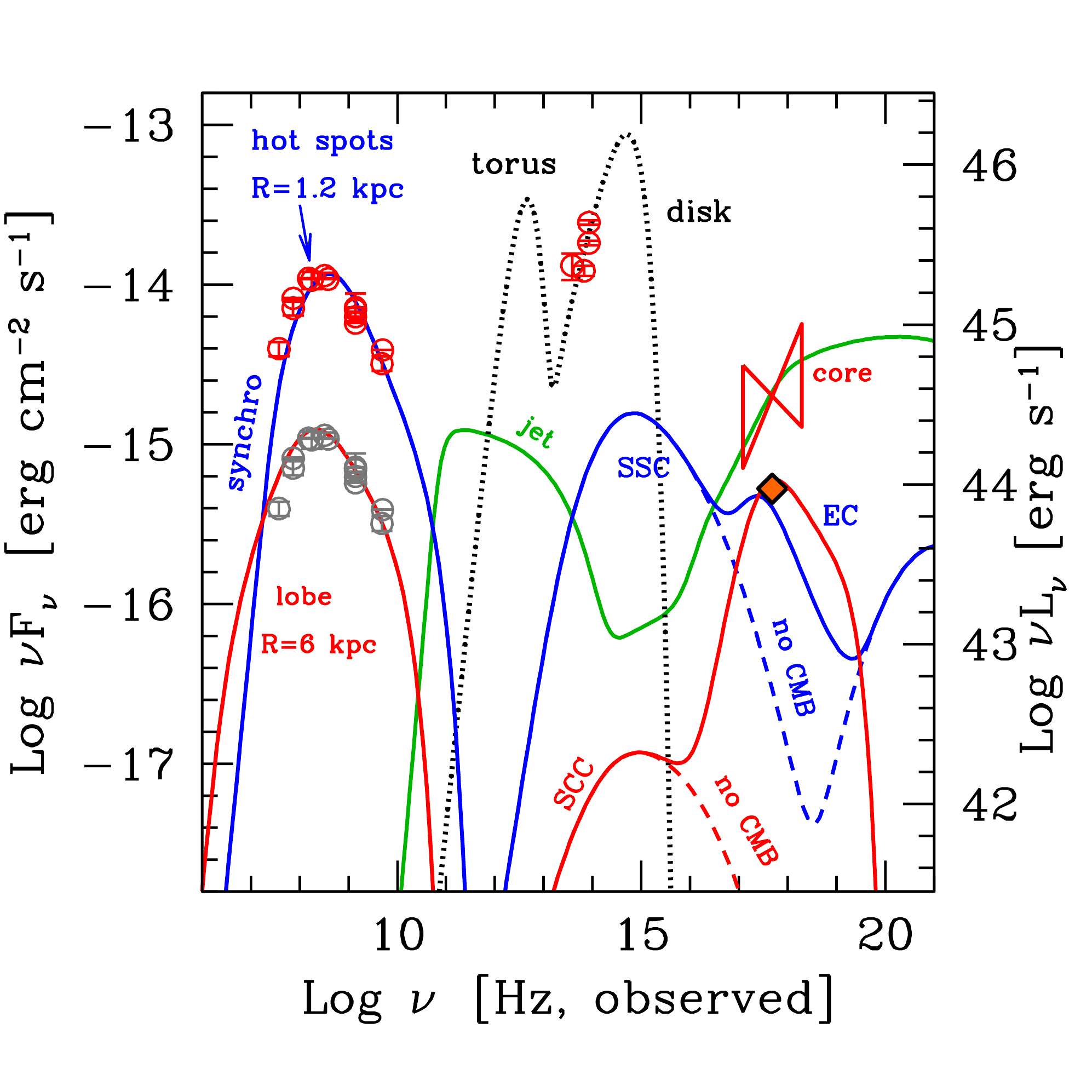}
    \caption{Reproducing the broadband SED of \so\ with either a jet$+$lobe or jet$+$hotspot emission model (see \S\ref{sec:model} for a detailed description of each). The contribution from the accretion disc and absorbing torus is marked by a dotted black line. The relativistic jet emission, shown as a solid green line, accounts for the measured core X-ray spectrum (red butterfly). The diffuse X-ray emission detected by \cxo\ (orange diamond) is fit with two models: a ``pure hotspot" model, shown as a solid blue line, where the X-rays arise entirely from compact, 1.2 kpc-radius hotspots, which are also responsible for the measured radio spectrum (marked by red open circles); and a ``pure-lobe'' model, shown as a solid red line, where the extended X-ray emission arises from 6 kpc-radius lobes, along with 15\%~of the radio emission\textsuperscript{\ref{foot:spec}} (open grey circles). Shown as dotted blue/red lines, and labeled ``no CMB'', is the result of the same models but where IC/CMB is switched off.    \label{fig:model}}
\end{figure}%
\begin{table*} 
\centering
\begin{tabular}{l l l l l l l l l l l l }
\hline
\hline
$L_{\rm d}$ &$R_{\rm diss}$ & $P^\prime_{\rm e}$  & $B$ & $\Gamma$ & $\theta_{\rm V}$ & $\gamma_{\rm b}$ & $\gamma_{\rm max}$ & $s_1$ & $s_2$  & $P_{\rm jet}$ \\ 
~[1] & [2] & [3] & [4] & [5] & [6] & [7] & [8] & [9] & [10] & [11]  \\
\hline   
$15.6$  & 288 & 43.8 & 2.3 & 10 & 10 & 50 & $3000$ & 1.5 & 2.2 & 47.8  \\ 
\hline
\hline 
\end{tabular}
\vskip 0.4 true cm
\caption{ Jet model parameters.
Col. [1]: disc luminosity, in  $10^{45}$ erg s$^{-1}$;
Col. [2]: distance of the dissipation region from the black hole, in units of $10^{15}$ cm;
Col. [3]: logarithm of the power injected in the form of relativistic electrons, in the co-moving frame, in erg s$^{-1}$;
Col. [4]: magnetic field, in G;  
Col. [5]: bulk Lorentz factor;
Col. [6]: viewing angle, in degrees;
Col. [7] and Col. [8]: break and maximum Lorentz factor of the injected electron distribution;
Col. [9] and Col. [10]: slopes of the injected electron distribution; 
Col. [11]: logarithm of the total kinetic plus magnetic jet power, in erg s$^{-1}$.
Quoted power and energy values refer to {\it one} jet. \label{tab:jet}}
\end{table*}
\begin{table*} 
\centering
\begin{tabular}{l l l l l l l l l l l l l l l l l}
\hline
\hline
Model &$R$ &$P_{\rm e}$ &$B$  &$\gamma_{\rm b}$ &$\gamma_{\rm max}$ &$s_1$ &$s_2$ 
&$E_{\rm e}$  &$E_{\rm B}$ &$U_{\rm B}/U_{\rm rad}$ &$U_{\rm B}/U_{\rm CMB}$  \\ 
~ &[1] &[2] &[3] &[4] &[5] &[6] &[7]&[8] &[9] &[10] &[11] \\
\hline   
lobe     & 5.8  & 46.2  & 0.12 & $10^3$  & $10^4$  &--1 & 4.0 & 58.1  & 58.1  & 2.4 & 1.8  \\  
hotspot  & 1.2  & 46.9  & 0.45 & 800   & $10^4$  &--1 & 4.5 & 58.1  & 57.2  & 4.4 & 26 \\
\hline
\hline 
\end{tabular}
\vskip 0.4 true cm
\caption{
Adopted parameters for the extended emission model; the pure lobe model corresponds to the solid red line in Figure \ref{fig:model}, whereas the pure hotpots is shown as a solid blue line.
Col. [1]: Region radius, in kpc;
Col. [2]: logarithm of the power injected in relativistic electrons in erg s$^{-1}$;
Col. [3]: magnetic field in mG;
Col. [4] and Col. [5]: break and maximum Lorenz factor of the injected electron distribution;
Col. [6] and Col. [7]: slopes of the injected electron distribution; 
Col. [8]: logarithm of the total energy in relativistic electrons, in erg;
Col. [9]: logarithm of the total energy in magnetic field, in erg.
Col. [10]: ratio of magnetic to radiation energy density, where the latter includes the CMB, synchrotron radiation (the disc and torus contributions are negligible). 
Col. [11]: ratio of magnetic to CMB radiation energy density. 
Energetics value refer to one hotspot/lobe, while the lobe flux shown in the figures corresponds to {two} hotspots/lobes.\label{tab:extended}}
\end{table*}
We follow closely \citet{ghisellini15} in modelling the hotspot and lobe radiation. The particle energy distribution function obeys equation \ref{eq:qgamma}, but with different parameters from the jets'. The size, combined with the injected power, constrain the magnetic field and particle energy density. 
We approximate each component as a sphere which is homogeneously filled with magnetic field\footnote{The lobe size is smaller than the field coherence length-scale, $\lambda=10$ kpc \citep{carilli+taylor02,celotti+fabian04}, consistent with the hypothesis that the magnetic field is indeed homogeneous.}.
Unlike for the hotspots, the magnetic field energy within the lobes is assumed to be nearly in equipartition with the electron energy ($E_{\rm e}\simeq E_{\rm B}$), which enables us to estimate the lobes' $B$-field strength. At the hotspot location, the lobes are injected with energy at rate $P_{\rm e}$; the electron distribution function is given again by Equation \ref{eq:qgamma}, but with a different set of parameters than for the jet and the hotspots. 
%
%
A complete list of the adopted parameters for the lobe and hotspot models is given in Table \ref{tab:extended}. Worth noting are the inferred $B$-field strength and the magnetic to CMB radiation energy density ratio: for the pure lobe model, $B=0.12$ mG and \ubrad$\simeq 2$, to be compared against $B=0.45$ mG and \ubrad$=28$ for the pure hotspot model. 
\subsection{Modelling results}
Figure \ref{fig:model} illustrates our best attempt at reproducing the broadband SED of \so\ with the models described in \S\ref{sec:model}. The black dotted line represents the contribution from the accretion disc and the dusty torus. 
The solid green line is the emission from the relativistic jet, which is responsible for the measured core X-rays.  The diffuse X-ray spectrum is fit by either a pure hotspot model (I., solid blue line), where the hotspots also produce the measured radio spectrum, or a pure lobe model (II., solid red line), where the lobes are responsible for 15\%~of the measured radio emission\footnote{For clarity, the model attempts to reproduce the same radio SED as measured (open red circles), but with an 85\% lower normalization (this artificial SED is shown as open grey circles). While this representation of the lobe spectrum must be correct below the peak frequency ($\simlt 8$ GHz observed-frame), where the very same electrons are responsible for both emission components, it is almost certainly not correct above the peak, where the emission is generated by higher energy electrons, with shorter synchrotron life times; on the other hand, this population would have measurable effects only at very high X-ray energies, where the SED is completely unconstrained. \label{foot:spec}}.
Assuming a pure hotspot model naturally requires a high degree of magnetization within the hotspots, with $U_{\rm B}/U_{\rm rad}\simeq 30$. As expected, the high $U_{\rm B}$ in the hotspots enhances the SSC contribution, and diminishes the relative role of IC/CMB in cooling the electrons. 
In contrast, the pure lobe model, which only accounts for 15\%~of the measured radio emission, yields $U_{\rm B}/U_{\rm rad}\simeq 2$ and $E_{\rm e}/ E_{\rm B}\simeq 1$, implying that IC/CMB dominates dramatically over SSC.
In both cases, the striking effect of IC/CMB is illustrated by the dashed blue/red lines (``no CMB''), where CMB/IC is switched off.
\section{Discussion}\label{sec:discussion}
The \cxo\ observations presented here have revealed that the X-ray counterpart to the $z=4.26$ radio galaxy \so\ is made of a compact core plus extended SE-to-NW emission, which accounts for $\simeq$30\%~of the flux and is aligned with the known hotspots seen at 5 GHz. The (loosely constrained) separation and centroid positions of the SW and NE X-ray sources allow for both a hotspot as well as a truly diffuse, lobe-like interpretation.  

When we attempt to reproduce the system SED with a jet model which accounts for the contribution from the either the jet termination hotspots (I.), or the jet-powered lobes (II.), including IC scattering off of CMB photons, we find that both scenarios require a high degree of magnetization, with supra-equipartition magnetic field energy densities. The magnetic to CMB radiation energy density ratio is even more extreme for the pure hotspot model (\ubrad$\simeq 25$, vs. $\simeq 3$ for the pure lobe case); this implies a lesser (relative) IC/CMB contribution with respect to SSC compared to the pure lobe scenario.

Although we considered admittedly extreme either-or scenarios, the results from this modelling exercise are not incompatible with the conclusions drawn by \citet{wu17}, who followed the same approach in modelling the SED of 4C~41.17 and 4C~03.24, i.e. the two other (known) $z > 3.5$ radio galaxies with (reliably) extended X-ray emission. For both systems, the energy density in magnetic field was found to vastly exceed that of the CMB radiation in the hotspots (whereas the opposite is true for the extended lobes). This implies that, in both systems, most of the X-ray signal is likely produced within the lobes, via IC/CMB, whereas most of the radio signal is produced within the hotspots, via synchrotron, thus washing out the steep $(1+z)^4$ enhancement of the X-ray vs. radio luminosity expected of IC/CMB. It is likely, albeit not proven, that a similar process takes place in \so: modulo the lack of a firm detection of truly extended lobes in this system, the (observed-frame) 5 GHz emission is clearly dominated by hotspots, where the magnetic energy density largely dominates over radiation. 

Admittedly, for \so\ to exhibit a similar behaviour as 4C~41.17 and 4C~03.24 we would require the (majority of the) off-nuclear X-ray emission detected by \cxo\ to be lobe-like, i.e. truly diffuse in nature. 
In 4C~41.17 and 4C~03.24, it was possible to decompose the X-rays into core, lobes, and hotspots. This is not the case in \so, which motivates our two bracketing cases. Nevertheless, just as in the case of. 4C 41.17 and 4C 03.24, the magnetic energy density is found to be substantially higher than the radiation energy density for either models. \\ 

The next question we wish to address is whether these new results have any meaningful implication for the larger population of high-$z$, jetted AGN, and in particular for the viability of the so-called CMB-quenching mechanism, which posits that the inferred dearth of high-$z$ radio galaxies \citep{volonteri11,sbarrato15} stems from a cooling offset which makes this population emit primarily in the X-rays rather than radio band. 

Under the assumptions discussed above, our modelling results paint a picture where, as expected, these objects are highly magnetized. 
A useful comparison can be drawn with the lower redshift ($z\simlt 2$) radio-loud sample examined by \citet{hardcastle04}: there, the quoted equipartition magnetic energy density values for the hotspots range between $10^{-12}$ and $10^{-9}$ J/m$^3$. Adopting $10^{-10.5}$ J/m$^3$ as an indicative value, and for a hotspot radius of 0.7 kpc, yields a total magnetic energy of about $10^{55}$ erg, which is indeed 2 to 4 orders of magnitude lower than the inferred values for the hotspots of the above three $z>3.5$ radio galaxies.   
Summarizing, radio galaxies at high-$z$ are already exceptional because they are clearly not radio-quenched; nevertheless the X-ray luminosities in \so, 4C~41.17, and 4C~03.24 are consistent with the expectation from highly magnetized lobes in a hotter CMB. In turn, this is consistent with the view that the CMB may quench less exceptional lobes. A broader theoretical investigation of the range of B-field, lobe sizes and intrinsic radio luminosity required for making \hz\ radio galaxies undetected by FIRST will be the subject of a forthcoming work. 

\noindent \textit{Data availability}: The main data underlying this article can be downloaded from the Chandra Data Archive, at \url{https://cda.harvard.edu/chaser/}, under Sequence Number 703172.

\bibliography{bibliography}
\bibliographystyle{mnras}

\end{document}